\documentclass{emulateapj}
\usepackage{float}

\newcommand{\appropto}{\mathrel{\vcenter{
  \offinterlineskip\halign{\hfil$##$\cr
    \propto\cr\noalign{\kern2pt}\sim\cr\noalign{\kern-2pt}}}}}

\begin{document}

\title{The Universal Relation of Galactic Chemical Evolution: The Origin of the Mass-Metallicity Relation}

\author{H. Jabran Zahid$^{1,3}$, Gabriel I. Dima$^{1}$, Rolf-Peter Kudritzki$^{1}$, Lisa J. Kewley$^{2}$, Margaret J. Geller$^{3}$, Ho Seong Hwang$^{3}$, John D. Silverman$^{4}$ \& Daichi Kashino$^{5}$ }
\affil{$^{1}$University of Hawaii at Manoa, Institute for Astronomy - 2680 Woodlawn Dr., Honolulu,  HI 96822, USA}
\affil{$^{2}$Australian National University, Research School of Astronomy and Astrophysics - Cotter Road, Weston Creek, ACT 2611, Australia}
\affil{$^{3}$Smithsonian Astrophysical Observatory - 60 Garden St., Cambridge, MA 02138, USA}
\affil{$^{4}$ University of Tokyo, Kavli Institute for the Physics and Mathematics of the Universe (WPI) - Kashiwanoha, Kashiwa, 277-8583, Japan}
\affil{$^{5}$Nagoya University, Division of Particle and Astrophysical Science - Nagoya, 464-8602, Japan
}

\begin{abstract}
We examine the mass-metallicity relation for $z\lesssim1.6$. The mass-metallicity relation follows a steep slope with a turnover or `knee' at stellar masses around $10^{10} M_\odot$. At stellar masses higher than the characteristic turnover mass, the mass-metallicity relation flattens as metallicities begin to saturate. We show that the redshift evolution of the mass-metallicity relation depends only on evolution of the characteristic turnover mass. The relationship between metallicity and the stellar mass normalized to the characteristic turnover mass is independent of redshift. We find that the redshift independent slope of the mass-metallicity relation is set by the slope of the relationship between gas mass and stellar mass. The turnover in the mass-metallicity relation occurs when the gas-phase oxygen abundance is high enough that the amount of oxygen locked up in low mass stars is an appreciable fraction of the amount of oxygen produced by massive stars. The characteristic turnover mass is the stellar mass where the stellar-to-gas mass ratio is unity. Numerical modeling suggests that the relationship between metallicity and stellar-to-gas mass ratio is a redshift independent, universal relationship followed by all galaxies as they evolve. The mass-metallicity relation originates from this more fundamental universal relationship between metallicity and stellar-to-gas mass ratio. We test the validity of this universal metallicity relation in local galaxies where stellar mass, metallicity and gas mass measurements are available. The data are consistent with a universal metallicity relation. We derive an equation for estimating the hydrogen gas mass from measurements of stellar mass and metallicity valid for $z\lesssim1.6$ and predict the cosmological evolution of galactic gas masses. 
\end{abstract}

\section{Introduction}


Gas flows and star formation govern the evolution of galaxies. A key diagnostic of gas flows and star formation in galaxies is the amount of heavy elements relative to hydrogen in the interstellar medium (ISM). Heavy elements are produced by massive stars and are dispersed into the ISM by stellar mass loss processes. Therefore, the ISM metal content is closely linked to the stellar mass of a galaxy. The heavy element abundance is measured relative to hydrogen and therefore also depends on the gas content of galaxies. Understanding the evolution of the gas-phase abundance in terms of gas flows and star formation is fundamental for developing a comprehensive theory of galaxy evolution.

Oxygen is the most abundant heavy element formed in the Universe. Therefore, the abundance of oxygen can be used as a proxy for the production of all heavy elements. The gas-phase oxygen abundance is correlated to the stellar mass in star-forming galaxies. This relation is known as the mass-metallicity (MZ) relation. The MZ relation was first observed in a small sample of nearby galaxies by \citet{Lequeux1979}. They showed that galaxy metallicity increases with stellar mass. Subsequently, \citet{Tremonti2004} measure the MZ relation of $\sim50,000$ star-forming galaxies in the local Universe. They find a tight MZ relation ($\sim$0.1 dex scatter) extending over three orders of magnitude in stellar mass. 

The MZ relation is one of the primary observations for measuring the chemical evolution of galaxies. In nearby galaxies, the MZ relation extends down to stellar masses of $\sim10^6 M_\odot$ \citep{Lee2006, Zahid2012a, Berg2012, Andrews2013}. A correlation between stellar mass and metallicity is observed not only in gas but also stars \citep{Gallazzi2005, Kudritzki2012, Kirby2013, Conroy2013a, Hosek2014}. The MZ relation is observed out to $z\sim3$ \citep[and many others]{Savaglio2005, Erb2006b, Maiolino2008, Mannucci2009, Zahid2011a, Yabe2012, Zahid2012b, Zahid2013b, Zahid2013e} and perhaps beyond \citep{Laskar2011, Moller2013}. The MZ relation also holds for individual star-forming regions within galaxies \citep{Rosales-Ortega2012}. Observations of the MZ relation reveal that the metallicities of galaxies increase with time. The MZ relation of the most massive galaxies flattens as the Universe evolves \citep{Zahid2013b}. This flattening is a result of an empirical upper limit in galaxy metallicity. The most massive galaxies, even at high redshifts, evolve chemically to this upper metallicity limit \citep{Zahid2013b, Zahid2013e}. The stellar mass where the MZ relation flattens is $\sim0.7$ dex lower now than at $z\sim0.8$ \citep{Zahid2013b}.

It is clear that the MZ relation depends on gas flows and star formation. Still, despite numerous studies of the MZ relation over the last few decades, the physical origin of the MZ relation remains uncertain. Metallicity is defined as the amount of oxygen \emph{relative} to hydrogen. Therefore, an increase in metallicity could result from star formation or metal-poor outflows. In the case of the latter, the metallicity of outflowing material is lower than the ISM metallicity leading to an increase in the amount of oxygen \emph{relative} to hydrogen. Conversely, metallicity can be reduced both by metal-rich outflows or inflows of metal-poor gas. Without clear observational constraints for inflows and outflows, the effects of gas flows and star formation remain degenerate. The MZ relation could possibly be the result of  metal-rich outflows \citep{Larson1974}, inefficient star formation in less massive galaxies \citep{Brooks2007, Finlator2008, Calura2009}, metal-poor inflows \citep{Dalcanton2004}, variations in the initial mass function \citep[IMF;][]{Koppen2007} or some combination of these physical processes. Uncovering the origin of the MZ relation is crucial for understanding gas flows and star formation in galaxies.

The scatter observed in the MZ relation is correlated with other physical properties. These correlations provide clues to the origin of the MZ relation. \citet{Ellison2008} first showed an anti-correlation between metallicity and specific star formation rate for galaxies at a fixed stellar mass. A relation between stellar mass, metallicity and star formation rate (SFR) is observed in local \citep{Mannucci2010, Lara-Lopez2010, Yates2012, Andrews2013} and high redshift galaxies \citep{Zahid2013e, Yabe2014, Troncoso2014}. At stellar masses $\lesssim10^{10.5}M_\odot$, galaxies with high SFRs typically have lower metallicities and vice versa. \citet{Mannucci2010} derive a relation between stellar mass, metallicity and SFR that minimizes the scatter of metallicity in the local galaxy population. They argue that the minimum scatter relation between stellar mass, metallicity and SFR that they derive is independent of redshift. They refer to this minimum scatter relation as the fundamental metallicity relation (FMR). For the FMR, the higher SFRs observed in high redshift galaxies account for their lower metallicities. However, the redshift independence of the relation between stellar mass, metallicity and SFR remains tentative \citep{Niino2012, Perez-Montero2013, Sanchez2013, Zahid2013e, Ly2014}. 

Both the SFR and metallicity are dependent on the gas content. The anti-correlation between metallicity and SFR is likely the result of variations in gas content \citep{Hughes2012, Lara-Lopez2013, Bothwell2013}. At a fixed stellar mass, galaxies with higher gas fractions will exhibit elevated SFRs and lower metallicity. Recently, \citet{Bothwell2013} present observations suggesting that the FMR derived by \citet{Mannucci2010} is the result of a  relation between stellar mass, metallicity and gas content. In the \citeauthor{Bothwell2013} interpretation, the SFR acts as a proxy for gas content in the FMR proposed by \citet{Mannucci2010}. \citet{Bothwell2013} are not able to investigate the redshift dependence of the relation between stellar mass, metallicity and gas content due to lack of measurements of {atomic} gas in galaxies outside the local Universe.

\begin{deluxetable}{cl}
\tablecaption{Key for Symbols}
\startdata
\hline
\hline
Symbol & Definition \\
\hline
\\
$M_\ast$ & stellar mass \\
$M_g$ & gas mass \\ 
$M_{HI}$ & neutral hydrogen mass \\
$M_{H2}$ & molecular hydrogen mass \\
$M_H$ & neutral + molecular hydrogen mass \\
$M_z$ & mass of oxygen \\
$M_g^o$ & mass of oxygen in gas-phase \\
$M_\ast^o$ & mass of oxygen in stars \\
$\Psi$ & star formation rate \\
$Z$ & mass density of oxygen relative to hydrogen $M_z/M_g$ \\
$R$ & return fraction \\
$Y$ & nucleosynthetic oxygen yield \\
$\zeta$ & oxygen mass loss factor \\
$Y_N$ & the net yield, $Y - \zeta$ \\ \\
\hline
& Fit Parameters \\
\hline
\\
$M_o$ & turnover mass\\
$Z_o$ & saturation metallicity \\
$\gamma$ & low mass end slope of MZ relation 
\enddata
\label{tab:key_sym}
\end{deluxetable}


Here we model the origin of the MZ relation, its evolution and the dependency of the scatter on SFR. We will show that the fundamental relationship of galactic chemical evolution is the relationship between metallicity and the stellar-to-gas mass ratio. We present the data and methods in Section 2 and 3, respectively. In Section 4 we derive the MZ relation for $z\lesssim1.6$. We show that the data are consistent with a single metallicity relation that is independent of redshift. In Section 5 and 6, we interpret our results by examining analytical and numerical models of chemical evolution, respectively. In Section 7 we show the observed relationship between metallicity and the stellar-to-gas mass ratio in local galaxies. We provide a discussion in Section 8 and a summary of our major results in Section 9. For reference, Table \ref{tab:key_sym} defines the symbols used in this work. We adopt the standard cosmology $(H_{0}, \Omega_{m}, \Omega_{\Lambda}) = (70$ km s$^{-1}$ Mpc$^{-1}$, 0.3, 0.7) and a \citet{Chabrier2003} IMF.

\section{Data}

We investigate the MZ relation for $z\lesssim1.6$ using data from the Sloan Digital Sky Survey (SDSS), Smithsonian Hectospec Lensing Survey (SHELS), Deep Extragalactic Evolutionary Probe 2 (DEEP2) and the FMOS-COSMOS survey. The MZ relations examined in this work are previously published in \citet{Zahid2013b, Zahid2013e}. In this section we describe the survey samples and selection criteria.

\subsection{SDSS Data}

We derive the local MZ relation using the SDSS DR7 main galaxy sample \citep{Abazajian2009}. The spectroscopic data consists of $\sim900,000$ galaxies spanning a redshift range of $0<z<0.7$. The survey has a limiting magnitude of $r = 17.8$ and covers 8000 deg$^2$ on the sky \citep{Strauss2002}. The nominal spectral range of the observations is $3800 - 9200\mathrm{\AA}$ with a spectral resolution of $R = 1800-2000$. {We adopt the line fluxes measured by the MPA/JHU\footnote{http://www.mpa-garching.mpg.de/SDSS/DR7/} and the $ugriz$-band c-model magnitudes.} The line fluxes are corrected for dust extinction using the \citet{Cardelli1989} extinction law assuming a case B recombination H$\alpha$/H$\beta$ value of 2.86 \citep{Hummer1987}. 

We derive metallicity from diagnostics based on the ratio of strong nebular emission lines. These strong-line metallicity diagnostics are calibrated under the assumption that stellar ionizing flux is powering nebular emission. The diagnostics are not calibrated to measure metallicities when active galactic nuclei (AGN) are significantly contributing to the ionizing flux. Fortunately, line flux ratios discriminate between star formation and AGN powered nebular emission. \citet[BPT]{Baldwin1981} first showed that [NII]$\lambda6584$/H$\alpha$ vs [OIII]$\lambda5007$/H$\beta$ diagram could be used to classify galaxies as star-forming or AGN. In order to derive a sample of star-forming galaxies we use the recent classification of \citet{Kewley2006}. Galaxies with
\begin{equation}
\mathrm{log([OIII]/H}\beta) > 0.61/(\mathrm{log([NII]/H}\alpha - 0.05) + 1.3 
\end{equation}
are classified as AGN and are removed from the sample.

Our analysis is based on the emission line fluxes of [OII]$\lambda3727,3729$, H$\beta$, [OIII]$\lambda5007$, H$\alpha$ and [NII]$\lambda6584$. \citet{Foster2012} show that signal-to-noise (S/N) cuts applied to the [OIII]$\lambda5007$ lines bias the measured MZ relation. This is because high metallicity galaxies typically have very weak [OIII]$\lambda5007$ emission. We apply no $S/N$ selection on the [OIII]$\lambda5007$ emission line. We require a S/N$>3$ in the line flux measurements of [OII]$\lambda3727,3729$, H$\beta$, H$\alpha$ and [NII]$\lambda6584$. We apply a lower redshift limit of $z>0.02$ to ensure that the $[OII]\lambda3727$ is redshifted into the nominal spectral range of the survey. {The SDSS sample spans a broad range of redshifts ($0<z<0.7$). An upper redshift limit of $z < 0.12$ is applied to minimize evolutionary effects}. The 3 arc-second spectroscopic fiber aperture only covers a fraction of the total galaxy flux. We derive the covering fraction by comparing the 3 arc-second fiber flux to the total flux. In order to avoid systematic aperture bias, we require an aperture covering fraction $>20\%$ as recommended by \citet{Kewley2005}. The local selected sample from the SDSS consists of $\sim51,000$ galaxies. 

\subsection{SHELS Data}

The SHELS survey \citep{Geller2005} consists of $\sim25,000$ galaxies ranging from $0<z<0.8$ in the F1 (Hwang et al., in preparation) and F2 \citep{Geller2014} fields of the Deep Lens Survey \citep{Wittman2002}. The two fields combined cover 8 deg$^2$ and are observed down to a limiting magnitude of $R=20.6$. The spectra are taken with Hectospec \citep{Fabricant2005}, a 300 fiber spectrograph mounted on the 6.5 m MMT. The nominal spectral range of the observations is $3700-9150\AA$ with a spectral resolution of $R\sim3000$. The $ugriz$-band c-model photometry is from the SDSS DR8 \citep{Padmanabhan2008}. 

We select galaxies in the SHELS sample by applying selection criteria similar to the criteria applied to the SDSS sample. AGN are removed from the sample using the \citet{Kewley2006} classification. We require a S/N$>3$ for [OII]$\lambda3727,3729$, H$\beta$, H$\alpha$ and [NII]$\lambda6584$ equivalent widths. {Applying our S/N selection to the observed line fluxes rather than equivalent widths yields a consistent sample.} The redshift range of the data is restricted to $0.2 < z < 0.38$. The lower redshift limit is set to sufficiently capture evolution between the SDSS and SHELS sample and the upper redshift limit is set by H$\alpha$ redshifting outside the nominal spectral range. In the restricted redshift range of our metallicity analysis, covering fraction is $>20\%$ and aperture bias is not an issue. The SHELS sample we select consists of $3577$ galaxies. 

\subsection{DEEP2 Data}

The DEEP2 survey \citep{Davis2003} consists of $50,000$ galaxies observed in the redshift range of $0.7<z<1.4$. We use data from the third data release\footnote{http://deep.ps.uci.edu/DR3/}. The data were observed in four fields covering 3.5 deg$^2$ down to a limiting magnitude of $R=24.1$. The spectra were taken with DEIMOS \citep{Faber2003} mounted on the 10 m Keck Telescope. The nominal spectral range is $6500-9100\AA$ observed at a resolution of $R\sim5000$. The $BRI-$band photometry was measured from images taken with the CFH12K camera on the 3.6 m Canada-France-Hawaii Telescope \citep{Coil2004}. For about half the galaxies in the sample, we have $K_s-$band photometry from the Wide Field Infrared Camera on the 5 m Hale Telescope \citep{Bundy2006}. Details of the emission line equivalent width measurements can be found in \citet{Zahid2011a}.

The [OII]$\lambda3727,3729$, H$\beta$ and [OIII]$\lambda5007$ emission lines required for our metallicity analysis are only observed in the redshift range of $0.75<z<0.82$. We require a S/N$>3$ in the [OII]$\lambda3727,3729$ and H$\beta$ emission line equivalent width measurements. Given the nominal spectral range of the data and the redshift of our sample, the H$\alpha$ and [NII]$\lambda6584$ emission lines are not observed and we are not able to apply the BPT classification. AGN contamination in the emission line galaxy sample is estimated to be small \citep{Weiner2007}. We identify and remove 17 AGN from the sample using X-ray observations \citep{Goulding2012}. The DEEP2 sample we select consists of 1254 galaxies.

\subsection{FMOS-COSMOS Data}

The FMOS-COSMOS survey (Silverman et al., in prep) is an ongoing near-infrared spectroscopic survey of star-forming galaxies in the redshift range of $1.4<z<1.7$ found in the central square degree of the COSMOS field \citep{Scoville2007}. The observations are carried out using FMOS \citep{Kimura2010} mounted on the 8 m Subaru Telescope. Galaxies are observed in the $H-$band (1.6 -  1.80 $\mu$m) in high resolution mode with spectral resolution of $R\sim2200$. A subsample of galaxies are also observed in $J-$band (1.11 - 1.35 $\mu$m). In the target redshift range, [NII]$\lambda6584$ and H$\alpha$ are observed in the $H-$band and [OIII]$\lambda5007$ and H$\beta$ are observed in the $J-$band. The COSMOS field has 30 bands of photometry covering UV to IR \citep{Ilbert2009}.

In order to ensure efficient detection, star-forming galaxies are pre-selected in the target redshift range using robust photometric redshifts \citep{Ilbert2009}, an sBzK color selection \citet{Daddi2004} and a $K_s-$band limiting magnitude of $K_s<23$. Each FMOS pointing is observed for 5 hours yielding a 3$\sigma$ line detection limit of $4 \times 10^{-17}$ ergs s$^{-1}$ cm$^{-2}$ corresponding to an unobscured SFR limit of $\sim5 M_\odot$ yr$^{-1}$. In all galaxies at least one emission line is detected with $>3\sigma$ significance in each individual galaxy spectrum. We identify the strongest observed emission line in our $H-$band observations as H$\alpha$. 

We derive metallicities using the [NII]$\lambda6584$/H$\alpha$ line flux ratio. Given our sensitivity limit and the typical metallicities and SFRs of galaxies at $z\sim1.6$, [NII]$\lambda6584$ is only detected in a fraction of individual galaxies. To derive an unbiased MZ relation, the spectra are summed in bins of stellar mass before line flux measurements are made. Each summed spectrum is derived from 15 or 16 individual galaxy spectra. Details of the metallicity analysis are presented in \citet{Zahid2013e}.

\subsection{Gas Mass Data}

The HI gas mass measurements are taken from the Arecibo Fast Alfa Survey \citep[ALFALFA;][]{Haynes2011}. We use the publicly available data presented in the $\alpha.$40 release\footnote{http://egg.astro.cornell.edu/alfalfa/data/index.php}. The data consist of $15,855$ detections over a 2800 deg$^{2}$ field. Of the observed detections, $95\%$ are associated with extragalactic sources. Nearby galaxies have large uncertainties in their distances which directly impacts the HI mass estimate. We require that galaxies have measured distances $>10$ Mpc. Additionally, we select objects which are designated as ``Code 1" (HI sources) and have significant detections of the 21 cm line ($S/N > 3$). The HI mass sample is cross-matched to the SDSS DR7 sample by the ALFALFA team. The cross-matched sample we select consists of 6399 galaxies.

In order to measure metallicities using the SDSS spectroscopy we apply additional selection criteria to the cross-matched sample. We require a S/N$>3$ for the [OII]$\lambda3727,3729$, H$\beta$, H$\alpha$ and [NII]$\lambda6584$ emission line flux measurements. Because much of the sample consists of nearby galaxies and the sample is small we do not apply a fiber aperture covering fraction selection as is done for the selected SDSS sample described in Section 2.1. The final cross-matched sample consists of 2633 galaxies.

Atomic and molecular hydrogen is present in star-forming galaxies. Unfortunately, we do not have direct estimates of the molecular hydrogen content of galaxies in our cross-matched sample. Instead we account for molecular hydrogen using the observed scaling between atomic and molecular gas mass derived as part of the COLD GASS survey \citep{Saintonge2011} conducted using the IRAM 30 m telescope. \citet{Saintonge2011} measure the molecular-to-atomic gas mass ratio in 350 local galaxies that are part of the ALFALFA survey. Galaxies are observed until CO is detected or until an upper limit for the molecular gas mass to stellar mass ratio of 1.5\% is reached. We estimate H$_2$ masses from the \citet{Saintonge2011} relation derived from a subset of the data where CO emission was detected. This CO detected subset of galaxies is primarily comprised of star-forming galaxies. Thus, the molecular-to-atomic gas mass ratio derived from this subset is appropriate for our sample. \citet{Saintonge2011} derive the relation 
\begin{equation}
\mathrm{log}\left( \frac{M_{H2}}{M_{HI}} \right)  = 0.425 \left[\mathrm{log}(M_\ast/M_\odot) - 10.7\right] -0.387 
\label{eq:mh2}
\end{equation}
where $M_{H2}$ and $M_{HI}$ are the molecular and atomic hydrogen mass, respectively. To obtain an estimate of the total hydrogen gas mass of galaxies in our cross-matched sample, we add the molecular hydrogen mass derived from Equation \ref{eq:mh2} to the HI mass measurements from ALFALFA. For galaxies at $M_\ast \lesssim 10^{10} M_\odot$, the molecular-to-atomic hydrogen mass fraction is $\lesssim20\%$. For the most massive galaxies in the sample, the molecular-to-atomic hydrogen mass fraction is $\lesssim50\%$.

\section{Methods}

\subsection{Stellar Mass}

We measure stellar masses using the Le Phare\footnote{http://www.cfht.hawaii.edu/$\sim$arnouts/LEPHARE/cfht\_lephare/ lephare.html} code developed by Arnouts \& Ilbert. We synthesize synthetic magnitudes using stellar population synthesis models of \citet{Bruzual2003} and a \citet{Chabrier2003} IMF. The synthetic magnitudes are generated by varying stellar population parameters. The models have two metallicities and seven exponentially decreasing star formation models (SFR $\propto e^{-t/\tau}$) with $\tau = 0.1,0.3,1,2,3,5,10,15$ and $30$ Gyrs. We apply the extinction law of \citet{Calzetti2000} allowing E(B$-$V) to range from 0 to 0.6. The ages of the stellar population range from 0 to 13 Gyrs. Each synthetic SED generated is normalized to solar luminosity. The stellar mass is the scaling factor between the synthetic SED and the observed photometry. There is a distribution of stellar masses derived which depend on the different stellar population parameters adopted to generate the synthetic SEDs. We adopt the median of this distribution as our stellar mass estimate. 

{To assess the relative accuracy of our stellar mass estimates, we compare our stellar masses with the stellar masses derived by the MPA/JHU group. The stellar masses we derive using LePhare are $\sim0.12$ dex lower than the MPA/JHU mass estimates. Differing IMFs account for $\sim0.04$ dex of the difference (\citeauthor{Chabrier2003} compared to \citeauthor{Kroupa2001}). We estimate the relative accuracy of our stellar masses by the root mean square (RMS) of the difference between the two methods. The RMS of the difference is $0.11$ dex. The absolute uncertainty in stellar mass is $\sim0.3$ dex \citep{Conroy2013b}. }


\subsection{Gas-Phase Oxygen Abundance}

Flux ratios of collisionally excited lines to recombination lines are sensitive to temperature and therefore metallicity. Various metallicity diagnostics which use ratios of strong emission lines have been empirically and/or theoretically calibrated \citep[for a review see][]{Kewley2008}. Empirical diagnostics are typically calibrated to metallicities measured from temperature sensitive auroral lines \citep[e.g.,][]{Pettini2004}. Theoretical calibrations rely on detailed photoionization modeling to derive metallicity diagnostics \citep[e.g.,][]{Kewley2002a}. \citet{Kewley2008} show that there are systematic offsets in the absolute metallicities derived using different diagnostics. However, they find that both the empirically and theoretically calibrated metallicity diagnostics are relatively accurate. Our results only depend on relative accuracy in measurements of metallicity. The metallicity diagnostics we use deliver relative accuracy at the level required for this work.

For the SDSS, SHELS and DEEP2 samples we derive metallicities using the $R23$ strong line method calibrated by \citet[hereafter KK04]{Kobulnicky2004}. A major advantage of this method is that it explicitly solves and corrects for the ionization parameter which may evolve with redshift \citep{Kewley2013a}. The relevant ratios of measured emission line intensities are
\begin{equation}
R23 = \frac{\mathrm{[OII]}\lambda3727 + \mathrm{[OIII]}\lambda4959 + \mathrm{[OIII]}\lambda5007}{\mathrm{H}\beta}
\end{equation}
and 
\begin{equation}
O32 = \frac{ \mathrm{[OIII]}\lambda4959 + \mathrm{[OIII]}\lambda5007}{\mathrm{[OII]}\lambda3727}.
\end{equation}
The SDSS line fluxes are corrected for dust extinction. For the SHELS and DEEP2 data we use the line equivalent widths\footnote{Equivalent widths may be substituted for dust corrected line flux ratios when measuring metallicities \citep{Kobulnicky2003b, Zahid2011a, Moustakas2011}.}. We assume that the [OIII]$\lambda5007$ to [OIII]$\lambda4959$ intensity ratio is 3 \citep{Osterbrock1989} and adopt a value of 1.33 times the [OIII]$\lambda5007$ intensity when summing the [OIII]$\lambda5007$ and [OIII]$\lambda4959$ line fluxes or equivalent widths. The $R23$ method is sensitive to the ionization state of the gas and the $O32$ ratio is used to correct for variations. The $R23$ method has two metallicity branches. All galaxies in this study are sufficiently massive to be on the upper branch. The intrinsic uncertainty of an individual measurement is $\sim0.1$ dex \citep{Kobulnicky2004}. 

A disadvantage of using the $R23$ method is that the emission lines used in the diagnostic are separated by $\sim1300\AA$. For high redshift galaxies, observing such large baselines is very time intensive and not feasible in large samples. The $N2$ method requires only the line flux ratio of the [NII]$\lambda6584$ to H$\alpha$. Because the lines are closely spaced in wavelength, an extinction correction is not required. We apply the $N2$ method calibrated by \citet[hereafter PP04]{Pettini2004} to the summed-spectra emission line fluxes measured from the FMOS-COSMOS sample at $z\sim1.6$. The intrinsic uncertainty of an individual measurement is $0.18$ dex.

There is an absolute offset of $\sim0.3$ dex between the metallicities measured using empirically and theoretically calibrated diagnostics. By applying different diagnostics to the same $\sim30,000$ galaxies in the SDSS, \citet{Kewley2008} derive formulae to convert metallicities from one baseline diagnostic to another. We convert metallicities measured for the FMOS-COSMOS sample using the PP04 diagnostic to the KK04 diagnostic using the conversion formulae in \citet{Kewley2008}. 

At metallicities above solar, the $N2$ diagnostic saturates \citep{Baldwin1981, Kewley2002a}. The metallicity of the highest stellar mass bin of the FMOS-COSMOS sample suffers from this saturation. Since we have a limited number of objects with individual line detections in the FMOS-COSMOS sample and we have no independent estimate of the metallicity for the galaxies affected by saturation, we are not able to quantify and correct for the systematic underestimate. From analysis of local SDSS galaxies, we estimate that the systematic effect of $N2$ saturation results in a $\sim0.03$ dex underestimate of the metallicity. Rather than apply a 0.03 dex correction to the data, we take a more conservative approach and add 0.03 dex to the uncertainty of the metallicity measured in the highest stellar mass bin of the FMOS-COSMOS sample.

Throughout this work, the metallicity is given as a ratio of the number of oxygen atoms to hydrogen atoms and is quoted as 12 + log(O/H).

\section{The Mass-Metallicity Relation}

\begin{figure*}
\begin{center}
\includegraphics[width=2\columnwidth]{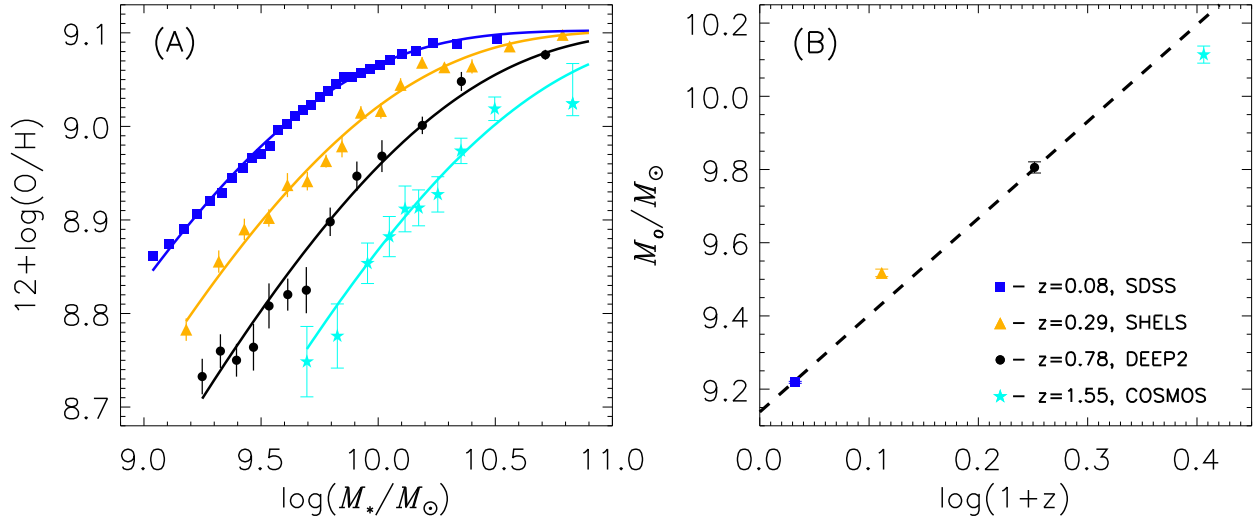}
\end{center}
\caption{(A) The MZ relation for $z\lesssim1.6$. The solid curves are the best single parameter model fits to the MZ relation. The model is defined in Equation \ref{eq:mzfit}. $Z_o$ and $\gamma$ fixed to the locally measured value from the SDSS data. (B) The best fit value of $M_o$ as a function of redshift. The dashed line is a fit to $M_o$ as a function of redshift and is given by Equation \ref{eq:mo}.}
\label{fig:mzr}
\end{figure*}

We derive the MZ relation for the SDSS, SHELS and DEEP2 samples by binning the metallicities measured in individual galaxies. We sort the SDSS, SHELS and DEEP2 data into 30, 16 and 13 equally populated stellar mass bins, respectively. We calculate the median stellar mass and metallicity in each bin. The metallicity error associated with the median in each bin is determined from bootstrapping the data. The MZ relation we measure from the FMOS-COSMOS sample is derived from emission lines measured from summed spectra sorted into 10 bins of stellar mass. The metallicity errors are calculated from propagating the observational uncertainties.

Many mathematical models may quantitatively fit data but not all models are physically interpretable. For example, the MZ relation can be fit by a polynomial. However, the best fit parameters derived from a polynomial fit can not be straightforwardly interpreted since the model parameters are not related to physical quantities. We propose a new model for fitting the MZ relation which we can physically interpret. This model is similar in form to the model proposed by \citet{Moustakas2011} and used in \citet{Zahid2013b}. Our interpretation of the fit parameters is presented in the following section (Section 5). We model the MZ relation as
\begin{equation}
12+\mathrm{log}(O/H) = Z_{o} + \mathrm{log} \left[ 1 - \mathrm{exp} \left( -\left[ \frac{M_\ast}{M_o} \right] ^{\gamma} \right) \right].
\label{eq:mzfit}
\end{equation}
In this model, $Z_o$ is the saturation metallicity. It quantifies the asymptotic upper metallicity limit \citep{Moustakas2011, Zahid2013b}. $M_o$ is the characteristic turnover mass above which the metallicity asymptotically approaches the upper metallicity limit, $Z_o$. At stellar masses $<M_o$, the MZ relation reduces to a power law with an index $\gamma$. {The primary difference between Equation \ref{eq:mzfit} and the model derived by \citet{Moustakas2011} is that an exponential rather than a power law appears in the argument of the logarithm}. 

\begin{deluxetable*}{cccccc}
\tablewidth{350pt}
\tablecaption{MZ Relation Fit}
\tablehead{\colhead{Sample} & \colhead{Redshift} &\colhead{$Z_o$} & \colhead{$\mathrm{log}(M_o/M_\odot)$} & \colhead{$\gamma$}  & \colhead{$\chi^{2}_{r}$}}
\startdata
\noalign{\smallskip}
\multicolumn{6}{c}{BEST FIT}\\
\hline
\noalign{\smallskip}
SDSS &0.08 & 9.102$\pm$0.002 & 9.219$\pm$0.007 & 0.513$\pm$0.009 & 1.89 \\
SHELS &0.29 & 9.102$\pm$0.004 & 9.52$\pm$0.02 & 0.52$\pm$0.02 & 1.68 \\
DEEP2 &0.78 & 9.10$\pm$0.01 & 9.80$\pm$0.05 & 0.52$\pm$0.04 & 1.28 \\
COSMOS &1.55 & 9.08$\pm$0.07 & 10.06$\pm$0.20 & 0.61$\pm$0.15 & 0.47 \\
\hline
\noalign{\smallskip}
\multicolumn{6}{c}{$Z_o$, $\gamma$ FIXED}\\
\noalign{\smallskip}
\hline
\noalign{\smallskip}
SDSS &0.08 & 9.102 & 9.219$\pm$0.003 & 0.513 & 1.89 \\
SHELS &0.29 & 9.102 & 9.52$\pm$0.01 & 0.513 & 1.46 \\
DEEP2 &0.78 & 9.102 & 9.81$\pm$0.02 & 0.513 & 1.08 \\
COSMOS &1.55 & 9.102 & 10.11$\pm$0.02 & 0.513 & 0.52 \\
\enddata
\label{tab:fit}
\end{deluxetable*}

We fit the MZ relation with the model parameterized by Equation \ref{eq:mzfit} using the \emph{MPFIT} package implemented in IDL \citep{Markwardt2009}. The data are inverse variance weighted and the errors are propagated through to the fit parameters. The best fit parameters and errors are given in the top half of Table \ref{tab:fit}. For all four samples, the values of $Z_o$ and $\gamma$ are consistent within the errors ($\lesssim2\sigma$). We also fit the data with a single parameter model in which we fix $Z_o$ and $\gamma$ to the SDSS values, allowing only $M_o$ to vary. The single parameter model values are listed in the bottom half of Table \ref{tab:fit}.

We plot the MZ relation for $z\lesssim1.6$ in Figure \ref{fig:mzr}A. The data show that at a fixed stellar mass, the metallicity increases as the Universe evolves. The MZ relation flattens at high stellar masses. Massive metal-rich galaxies at high redshifts have metallicities comparable to massive local galaxies. In \citet{Zahid2013b} we show the flattening of the MZ relation results from galaxies enriching to an empirical metallicity limit. This limit is parameterized by $Z_o$ and does not evolve significantly with redshift. The solid curves in Figure \ref{fig:mzr}A show the single parameter model fits. The redshift evolution of the MZ relation can be parameterized solely by evolution in the characteristic turnover mass. The value of $M_o$ derived from fitting the single parameter model is plotted in Figure \ref{fig:mzr}B. The single parameter model fit to the data shows that the characteristic turnover mass, $M_o$, is an order of magnitude larger at $z\sim1.6$. In Section 5 and 6 we elucidate the physical basis for this evolution. 


We test the null hypothesis that three parameter model better reproduces the data. The reduced $\chi^2$ values for the three and single parameter model fits are given in Table \ref{tab:fit}. An F-test analysis confirms that the three parameter model does not provide an statistically significant improvement to the fit.

\begin{figure*}
\begin{center}
\includegraphics[width=2\columnwidth]{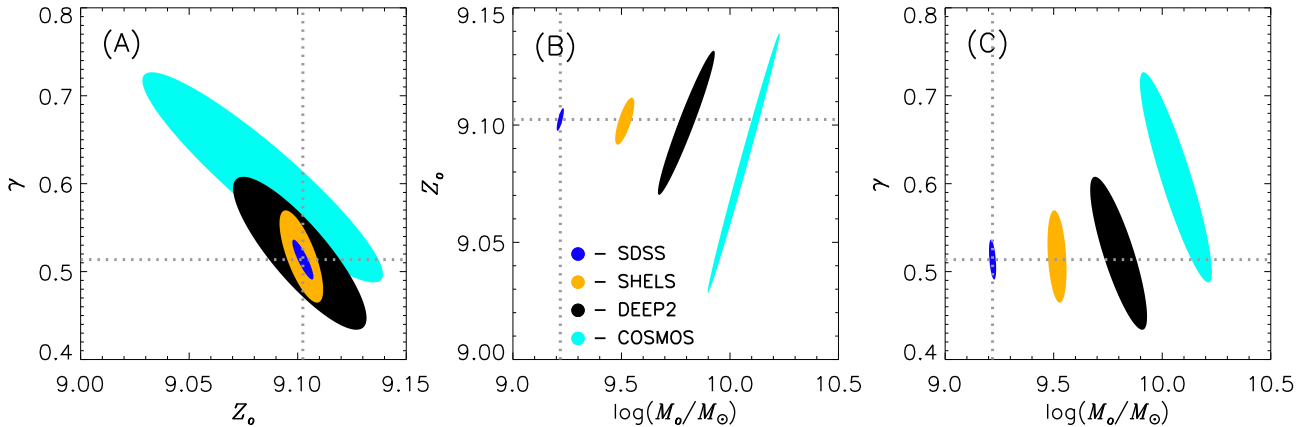}
\end{center}
\caption{The error ellipses indicating the covariance between the fit parameters. The best-fit values of $Z_o$, $M_o$ and $\gamma$ and their errors are given in the top half of Table \ref{tab:fit}. {For clarity, the error ellipses for the FMOS-COSMOS fit are decreased by a scale factor of three}. The parameter errors are determined by propagating the observational uncertainties and the error ellipses are calculated from the covariance matrix. The dotted gray lines indicate the best fit value of $Z_o$ and $\gamma$ from the SDSS data.}
\label{fig:ellipse}
\end{figure*}

The parameters of the model given by Equation \ref{eq:mzfit} are covariant. In Figure \ref{fig:ellipse} we examine the covariance of the best-fit parameters. We plot the 95\% confidence ellipse. For clarity in the figure, the error ellipses for the FMOS-COSMOS fit are decreased by a scale factor of three. The dotted gray lines are the best-fit values of the parameters for the SDSS data. In Figure \ref{fig:ellipse}A we plot the covariance of $\gamma$ and $Z_o$. Figure \ref{fig:ellipse}B and \ref{fig:ellipse}C show the covariance between $Z_o$ and $M_o$ and $\gamma$ and $M_o$, respectively. The data show that there is clear evolution in $M_o$. The orientation of the error ellipse indicates that $Z_o$ and $M_o$ are weakly covariant in all four samples. There is little to no covariance between $Z_o$ and $\gamma$ for the SDSS and SHELS samples. There is a small degree of covariance between $Z_o$ and $\gamma$ for the DEEP2 and FMOS-COSMOS sample. 

We conclude that the one parameter model is sufficient to describe the data and the best-fit value of $M_o$ is not strongly dependent on our choice to fix $Z_o$ and $\gamma$ to the SDSS values.

The MZ relation for $z\lesssim1.6$ is a simple function of redshift. The parameters in Equation \ref{eq:mzfit} that fit the redshift dependent MZ relation are
\begin{equation}
Z_o = 9.102 \pm 0.002,
\end{equation}
\begin{equation}
\gamma = 0.513 \pm 0.009,
\end{equation} 
and 
\begin{equation}
\mathrm{log}(M_o/M_\odot) = (9.138 \pm 0.003) + (2.64 \pm 0.05) \, \mathrm{log}(1+z).
\label{eq:mo}
\end{equation}
The redshift evolution of the MZ relation for $z\lesssim1.6$ is sufficiently parameterized by evolution in the characteristic turnover mass, $M_o$.

\begin{figure*}
\begin{center}
\includegraphics[width=1.5\columnwidth]{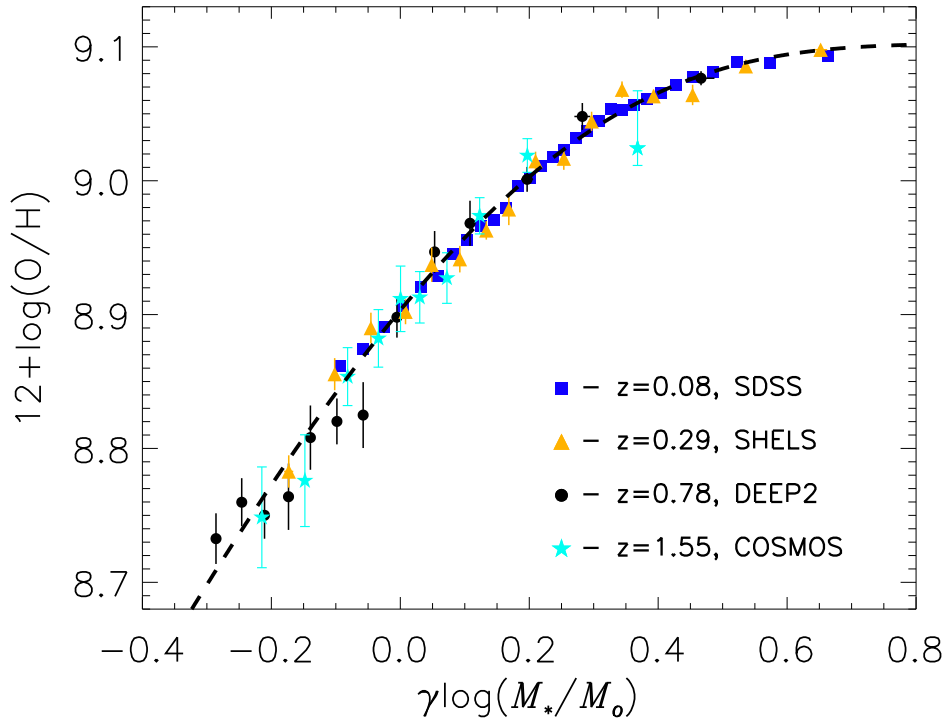}
\end{center}
\caption{The metallicity plotted against the stellar mass normalized to the measured characteristic turnover mass, $M_o$ of each data set. The dashed curve is given by Equation \ref{eq:mzfit} and the parameters are given in Table \ref{tab:fit} under the heading ``$Z_o, \gamma$ FIXED".}
\label{fig:zfg}
\end{figure*}

\emph{The data are consistent with a single, redshift independent value for $Z_o$ and $\gamma$.} The fact that the MZ relation evolution only depends on the characteristic turnover mass, $M_o$, means that relation between metallicity and stellar mass scaled by $M_o$ is independent of redshift. We plot the relation between metallicity and $\gamma \mathrm{log}(M_\ast/M_o)$ in Figure \ref{fig:zfg}. The key to unlocking the origin of the MZ relation is our ability to explain the relation plotted in Figure \ref{fig:zfg}. Thus we must determine physical meaning of $Z_o$, $\gamma$ and $M_o$.

\section{An Analytical Model of Chemical Evolution}

We begin our exploration of the physical origin of the MZ relation by examining analytical models of chemical evolution. The equations below do not necessarily have analytical solutions. We will make several physically motivated simplifying assumptions to arrive at an analytical solution. {The model we derive below is the inflow model first introduced by \citet{Larson1972}. The novel aspect of our model is our treatment of the impact of outflows.} 

While metallicities are traditionally quoted as a number density of oxygen to hydrogen, in the following discussion we define the metallicity as $Z \equiv M_z/M_g$. Here $M_z$ is the mass of oxygen in the gas-phase and $M_g$ is the hydrogen gas mass. A constant scale factor relates metallicity defined in terms of mass density to metallicity defined in terms of number density \citep{Peeples2011}.  We analyze the case of a single galaxy as it evolves chemically. We can thus parameterize the problem only in terms of stellar mass. To model the chemical evolution of arbitrary galaxy populations, time must be included in the equations. In this case, partial derivatives replace all derivatives below. The analytical solution that we derive based on a single galaxy is generalizable to the case of an arbitrary galaxy population. Our interpretation is independent of whether we analyze the case of a single galaxy or a population of galaxies.

We start with the equation of chemical evolution describing the change in the metallicity with respect to stellar mass. This is given by
\begin{equation}
\frac{\mathrm{d}Z}{\mathrm{d}M_\ast} = \frac{\mathrm{d}}{\mathrm{d}M_\ast} \left( \frac{M_z}{M_g} \right) = \frac{1}{M_g} \left( \frac{\mathrm{d}M_z}{\mathrm{d}M_\ast} - Z \frac{\mathrm{d}M_g}{\mathrm{d}M_\ast} \right).
\label{eq:dmzr}
\end{equation}
From observations we know that at low stellar masses we have an MZ relation with a positive power law index. In this case we necessarily have
\begin{equation}
\frac{\mathrm{d}M_z}{\mathrm{d}M_\ast} > Z \frac{\mathrm{d}M_g}{\mathrm{d}M_\ast}.
\label{eq:dzanal}
\end{equation}
When the metallicity is small, the second term of Equation \ref{eq:dmzr} ($Z \frac{\mathrm{d}M_g}{\mathrm{d}M_\ast}$) is negligible. From measurements of the stellar and gas content of galaxies in the local Universe, we have that $M_g \appropto \sqrt{M_\ast}$ \citep{Papastergis2012, Peeples2013}. The change in gas mass with respect to stellar mass is then $\mathrm{d}M_g/\mathrm{d}M_\ast \appropto 1/\sqrt{M_\ast}$. At higher stellar masses, the second term of the right hand side of Equation \ref{eq:dmzr} ($Z \frac{\mathrm{d}M_g}{\mathrm{d}M_\ast}$) tends to zero. In this case, the change in metallicity with respect to stellar mass can be approximated by
\begin{equation}
\frac{\mathrm{d}Z}{\mathrm{d}M_\ast} \approx \frac{1}{M_g} \frac{\mathrm{d}M_z}{\mathrm{d}M_\ast}.
\label{eq:dmzr_approx}
\end{equation}
This equation simply states that the chemical evolution of a galaxy is dominated by the production of metals and not by a slowly changing gas reservoir.

To solve Equation \ref{eq:dmzr_approx} we must define the right hand side. The change in oxygen mass is given by
\begin{equation}
\mathrm{d}M_z = Y \mathrm{d}M_\ast - Z\mathrm{d}M_\ast + RZ\mathrm{d}M_\ast + Z_{i} \mathrm{d}M_{i} - Z_{w} \mathrm{d}M_{w}.
\label{eq:dmz}
\end{equation}
The first term on the right hand side is the production of newly synthesized oxygen where $Y$ is the nucleosynthetic yield. $Y$ is the mass of oxygen created per unit SFR. d$M_\ast$ is the mass of newly formed stars, i.e. the SFR. $Y$ does not depend strongly on any galaxy properties \citep{Thomas1998, Kobayashi2006}. We treat $Y$ as a constant. The second term on the right hand side is the mass of oxygen in the ISM that forms into stars. The second term is negative indicating that the oxygen going into stars is taken out of the ISM. The third term on the right hand side is the mass of oxygen that goes into stars but is returned back to the ISM via stellar mass loss where $R$ is the fraction of mass returned back to the ISM. The third term represents previously synthesized oxygen returned back to the ISM by stellar winds and supernovae. The timescale for stellar mass return is short compared to the galaxy evolution timescale. We assume that stellar mass is instantaneously returned back into the ISM. The fourth and fifth terms represent the inflows and outflows (winds) of oxygen, respectively. The terms $Z_i$, d$M_i$, $Z_w$ and d$M_w$ are the inflow metallicity, mass rate of inflow, outflow metallicity and mass rate of outflows, respectively. Dividing Equation \ref{eq:dmz} by d$M_\ast$ we obtain
\begin{equation}
\frac{\mathrm{d}M_z}{\mathrm{d}M_\ast} = Y - Z(1-R) + Z_i \frac{\mathrm{d}M_i}{\mathrm{d}M_\ast} - Z_w \frac{\mathrm{d}M_w}{\mathrm{d}M_\ast}.
\end{equation}
This equation is the rate of change of the oxygen mass with respect to the SFR.

We combine the effects of outflows and inflows into a single factor defined as
\begin{equation}
\zeta \equiv  Z_w \frac{\mathrm{d}M_w}{\mathrm{d}M_\ast} - Z_i \frac{\mathrm{d}M_i}{\mathrm{d}M_\ast}.
\label{eq:zeta}
\end{equation}
$\zeta$ is the oxygen mass loss factor. Each of the terms on the right hand side of Equation \ref{eq:zeta} are uncertain \citep[see][]{Zahid2014a}. However, in \citet{Zahid2012b} we derive empirical constraints for the net loss of oxygen from galaxies. {We estimate the total mass of oxygen produced by galaxies from their current stellar mass. We estimate the mass of oxygen in the gas-phase from the MZ relation and the scaling between stellar mass and gas mass observed in local galaxies. We empirically constrain the mass of oxygen locked up in stars by constructing a self-consistent empirical model based on multi-epoch observations of SFRs and metallicities of galaxies.} Based on these estimates, we show that the total mass of oxygen expelled from galaxies over their lifetime is nearly proportional to their stellar mass. This means
\begin{equation}
\int_0^{M_\ast} \zeta \mathrm{d}M^{\prime}_\ast \propto M_\ast,
\end{equation}
implying that $\zeta$ is constant. In a more recent empirical analysis, \citet{Peeples2013} confirm that the mass of oxygen lost relative to the mass of oxygen produced is independent of stellar mass in star-forming galaxies. They estimate that on average, galaxies lose $\sim80\%$ of the oxygen they produce. The physical mechanism of outflows is not well understood though it is generally considered to be driven by energy/momentum from massive stars. Perhaps it is not that surprising that the mass of oxygen expelled from a galaxy scales with stellar mass since the total energy/momentum which is responsible for driving outflows in a galaxy is directly proportional to current stellar mass. With the adoption of $\zeta$ as a constant we have
\begin{equation}
\frac{\mathrm{d}Z}{\mathrm{d}M_\ast} \approx \frac{Y_N - Z(1-R)}{M_g}.
\label{eq:dmzr_final}
\end{equation}
Here we have combined the production and loss of oxygen, both of which are proportional to stellar mass, into a single term defined as $Y_N \equiv Y - \zeta$. We refer to this as the net yield. The net yield is the mass of oxygen produced by star formation modulo the oxygen expelled from the ISM. 

The measured hydrogen gas mass of star forming galaxies is reasonably well described by a power law over $\sim4$ decades in stellar mass \citep{Papastergis2012, Peeples2013}. To solve Equation \ref{eq:dmzr_final} we parameterize the relation between gas mass and stellar mass by 
\begin{equation}
M_g = GM_\ast^g, 
\label{eq:mgdef}
\end{equation}
where $G$ is the zero point and $g$ is the power law index of the relation. The solution to Equation \ref{eq:dmzr_final} is 
\begin{equation}
Z(M_\ast) = \frac{Y_N}{1-R} \left[ 1 - \mathrm{exp} \left( -\left[\frac{1-R}{1-g} \right] \frac{M_\ast}{M_g} \right) \right].
\label{eq:mzr}
\end{equation}
This equation has the same form as the equation we fit to the MZ relation (Equation \ref{eq:mzfit}). By taking the logarithm of Equation \ref{eq:mzr}, we can directly relate our fit parameters in Equation \ref{eq:mzfit} to the physical parameters analytically describing chemical evolution. The asymptotic metallicity $Z_o$ is
\begin {equation}
Z_o = \mathrm{log}\left(\frac{Y_N}{1-R}\right).
\end{equation}
The maximum metallicity observed in galaxies is set by the net yield, $Y_N$. From Equation \ref{eq:dmzr_final} we see that the metallicity saturates (d$Z$/d$M_\ast \approx 0$) when the amount of metals produced, $Y_N$, equals the amount of metals locked up in stars, $Z(1-R)$. The arguments of the exponentials in Equations \ref{eq:mzfit} and \ref{eq:mzr} can be equated to give
\begin {equation}
\frac{1-R}{1-g} \left( \frac{M_\ast}{M_g} \right) \approx \frac{M_\ast}{M_g} \approx  \left(\frac{M_\ast}{M_o}\right)^\gamma  .
\label{eq:momg}
\end{equation}
The return fraction $R$ and the power law index of the gas mass relation, $g$, are nearly equal ($R\sim g \sim 0.5$). Hereafter, we drop the prefactor term of the left hand side of Equation \ref{eq:momg}. Substituting our relation for the gas mass as a function of stellar mass from Equation \ref{eq:mgdef} we have
\begin{equation}
\left(\frac{M_\ast}{M_o}\right)^\gamma \approx \frac{M_\ast^{1-g}}{G}
\end{equation}
The low mass end slope we fit to the MZ relation, $\gamma$, is related to the power law index of the gas mass relation by $\gamma = 1 - g$. The characteristic turnover mass, $M_o$, is related to the zero point of the relation between gas mass and stellar mass by $M_o = G^{1/\gamma}$. $M_o$ is the stellar mass at which the gas-to-stellar mass ratio is unity. This interpretation suggests that the redshift evolution of the MZ relation is due to the larger gas masses of galaxies at early times. {In Figure \ref{fig:zfg} we show a redshift independent relation between metallicity and $(M_\ast/M_o)^{\gamma}$. From examination of Equation \ref{eq:momg}, we learn that $(M_\ast/M_o)^{\gamma} = M_\ast/M_g$. Thus, the redshift independent relation plotted in Figure \ref{fig:zfg} is a relation between metallicity and stellar-to-gas mass ratio.}

We test the consistency of our interpretation by comparing the observed gas-to-stellar mass relation in local galaxies with the SDSS MZ relation fit parameters. \citet{Peeples2013} derive the relationship between gas mass and stellar mass from $\sim260$ star-forming galaxies where both the atomic and molecular gas masses are measured. The relation they derive is given by $\mathrm{log}(M_g/M_\ast) = -0.48 \, \mathrm{log}(M_\ast/M_\odot) + 4.39$. For this relation, $g = 0.52$ and $M_g = M_\ast$ at log$(M_\ast/M_\odot) = 9.15$. Our interpretation says that the $\gamma = 1-g$ and $M_o$ is the stellar mass where the gas-to-stellar mass ratio is unity. Based on the gas-to-stellar mass relation measured by \citet{Peeples2013} we would predict an MZ relation with $\gamma = 1 - g = 0.48$ and $M_o = 9.15$. We measure $\gamma = 0.51$ and $M_o = 9.2$ for the local relation. The fit parameters of the SDSS MZ relation are remarkably consistent with the measured gas-to-stellar mass relation in local galaxies.

\section{A Numerical Model of Chemical Evolution}

The simplifying assumptions made in deriving an analytical solution may obfuscate our interpretation. In this section we numerically model the chemical evolution of galaxies. The model serves as a heuristic tool to explore the validity of our interpretation of the MZ relation. 

We self-consistently model the metallicity of galaxy populations as they evolve applying empirical constraints for their stellar mass growth. To derive the star formation history (SFH) of galaxies, we use the approach developed in \citet{Zahid2012b} (also see \citealt{Leitner2012}). Galaxies exhibit a tight relation ($\sim0.25$ dex scatter) between stellar mass and SFR out to at least $z\sim2$ \citep{Noeske2007a, Salim2007, Elbaz2007, Pannella2009, Whitaker2012, Zahid2012b, Kashino2013}. Observations allow us to parameterize the SFR as a function of stellar mass and redshift. To derive the stellar mass histories of star-forming galaxies, we require that as galaxies evolve, they remain on the mean stellar mass-SFR (MS) relation at all epochs. The stellar mass history is given by
\begin{equation}
M_\ast(t) = (1-R) \int_{t_i}^{t} \Psi(M_\ast, t^\prime) dt^\prime.
\label{eq:sfh}
\end{equation}
$\Psi(M_\ast, t)$ is the star formation rate as a function of stellar mass and time. It is derived from observations of the MS relation at several epochs and is given by Equation 13 in \citet{Zahid2012b}. The integration is carried out from some initial time, $t_i$ to some later time, $t$.

The rate at which oxygen accumulates in the ISM is given by 
\begin{equation}
\frac{\mathrm{d}M_g^o}{\mathrm{d}t} = Y_N \Psi -  \frac{\mathrm{d}M_\ast^o}{\mathrm{d}t}
\label{eq:mgo}
\end{equation}
Here $M_g^o$  is the mass of oxygen in the gas phase. The first term on the right hand side is the net production term where as before, $Y_N$, is the net yield. The first term accounts for oxygen production and loss in outflows. The second term on the right hand side is the rate at which oxygen is locked up into stars and is given by
\begin{equation}
\frac{\mathrm{d}M_\ast^o}{\mathrm{d}t} = (1-R) Z \Psi.
\end{equation}
Here $M_\ast^o$ is the mass of oxygen locked up in stars. Both $Z$ and $\Psi$ are explicitly dependent on time and stellar mass. We set the return fraction to $R=0.45$ which is appropriate for a Chabrier IMF \citep{Leitner2011}. 

The time rate of change of the metallicity depends on the mass of oxygen produced, the mass of oxygen locked up in stars, which is itself dependent on the metallicity of the gas at the time of star formation and the change in the gas content as the galaxy evolves. The time rate of change of metallicity is given by
\begin{equation}
\frac{\mathrm{d}Z}{\mathrm{d}t} = \frac{\mathrm{d}}{\mathrm{d}t}\left( \frac{M_g^o}{M_g} \right), 
\label{eq:dznum}
\end{equation}
Equation \ref{eq:dznum} is just a restatement of Equation \ref{eq:dmzr}. We can numerically solve Equation \ref{eq:dznum} without the simplifying assumptions required in arriving at an analytical solution. The last quantity we need to numerically solve Equation \ref{eq:dznum} is the gas mass which we parameterize as a function of stellar mass and time:
\begin{equation}
\mathrm{log}( M_g/M_\odot) = G + \alpha \, \mathrm{log}(1+z) + \beta \, \mathrm{log}(M_\ast/M_\odot).
\label{eq:gmh}
\end{equation}
In this parameterization $G$ is the zero point of the gas mass at $z=0$ and $\alpha$ and $\beta$ parameterize the redshift and stellar mass dependence, respectively.

The solution to Equation \ref{eq:dznum} is the metallicity history of an individual galaxy as it evolves. Equations \ref{eq:mgo} - \ref{eq:dznum} are three coupled differential equations. We numerically solve these equations using an iterative method and a time step of 0.01 Myr. We solve Equation \ref{eq:dznum} for a population of galaxies which covers a wide range in stellar mass. From the metallicity history of a population of galaxies, we can calculate the MZ relation at each observed epoch. $Y_N$, $G$, $\alpha$ and $\beta$ are free parameters in our model. We vary these parameters in order to reproduce the observed MZ relations.

\begin{figure}
\begin{center}
\includegraphics[width=\columnwidth]{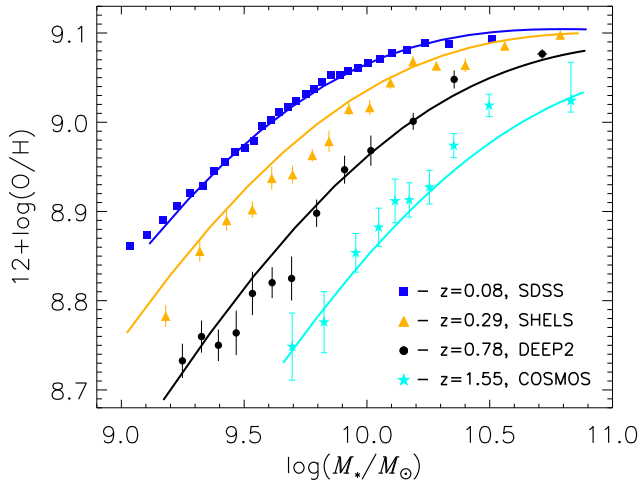}
\end{center}
\caption{The MZ relation ranging for $z\lesssim1.6$. The solid curves are the MZ relations determined from the numerical model by solving Equation \ref{eq:dznum}.}
\label{fig:model_fit}
\end{figure}

{In Figure \ref{fig:model_fit} we show the model MZ relations plotted over the observed MZ relations. We emphasize that our numerical model reproduces the data well considering that both the metallicities and stellar masses carry $\sim0.3$ dex absolute systematic uncertainties \citep{Kewley2008, Conroy2013b}. The net yield in our model is $Y_N = 5.7\times10^{-3}$ and the gas mass relation in our model is given by 
\begin{equation}
\mathrm{log}( M_g/M_\odot) = 5.04 + 1.34 \, \mathrm{log}(1+z) + 0.44 \, \mathrm{log}(M_\ast/M_\odot).
\label{eq:gm_model}
\end{equation} 
The power law index of the model relation between stellar mass and gas mass that reproduces the observed MZ relations is $\beta = 0.44$. Based on our interpretation of the MZ relation fit parameters, this implies a slope of the MZ relation $\gamma = 1- \beta = 0.56$. This value of $\gamma$ is close to our measured value of $\gamma = 0.513$.}

{The gas mass relation given by Equation \ref{eq:gm_model} is tuned to reproduce our observations of the MZ relation. This model gas mass relation is completely independent of the parameters we fit to the  MZ relation in Section 4 (see Table \ref{tab:fit}). Based on our interpretation of the MZ relation fit parameters presented in Section 5, we can derive the gas mass as a function of stellar mass and redshift from the MZ relation fit parameters we measure. Thus, a comparison of the model gas mass relation given by Equation \ref{eq:gm_model} with the relation implied by our observations provides an independent consistency check of our \emph{interpretation} of the MZ relation fit parameters. To derive the gas mass relation from the MZ relation fit parameters, we solve Equation \ref{eq:momg} for gas mass giving
\begin{equation}
M_g = M_o^\gamma \, M_\ast^{1-\gamma}.
\end{equation}
Inputting our measured values of $\gamma$ (= 0.51) and $M_o$ (Equation \ref{eq:mo}) gives
\begin{equation}
\mathrm{log}( M_g/M_\odot) = 4.69 + 1.35 \, \mathrm{log}(1+z) + 0.49 \, \mathrm{log}(M_\ast/M_\odot).
\label{eq:gm_obs}
\end{equation} 
The gas mass relation derived from our observations (Equation \ref{eq:gm_obs}) is very similar to the relation required in our model to reproduce the MZ relation (Equation \ref{eq:gm_model}).}  

The purpose of this exercise is to validate many of the assumptions made in the previous section {and to test our interpretation}. The model parameters for the gas mass relation are derived by fitting to the observed MZ relations using the simple equations of chemical evolution given by Equations \ref{eq:mgo} - \ref{eq:dznum}. The analysis based on our models is independent of the analysis presented in Section 5. The fact that we are able to reproduce the evolution of the MZ relation suggests that the interpretation presented in the previous section is supported by the observed MZ relations. In particular, the model analysis confirms that the following simplifications and interpretations are consistent with the data: 1) The build-up of metals, and not a changing gas reservoir, is the dominant process governing the evolution of the metallicity in individual galaxies. The first term on the right hand side of Equation \ref{eq:dmzr} is the dominant term. 2) The slope of the MZ relation is related to the slope of the gas mass relation and is reasonably approximated by $\gamma \approx 1 - g$, where $g$ is the power law index of the relation between gas mass and stellar mass. Thus, the slope of the gas-to-stellar mass relation is not strongly dependent on redshift. 3) The evolution of the characteristic turnover mass of the MZ relation, $M_o$, is related to the evolution of the zero point of the gas-to-stellar mass relation. 4) The MZ relation saturates when the mass of oxygen produced by massive stars equals the mass of oxygen locked up by low mass stars.

\begin{figure*}
\begin{center}
\includegraphics[width=2\columnwidth]{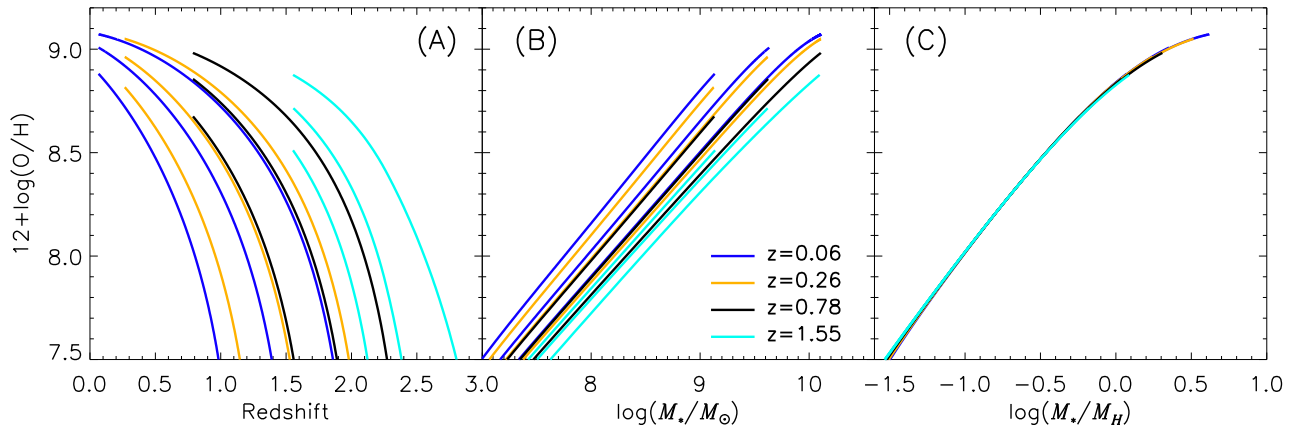}
\end{center}
\caption{The metallicity as a function of (A) redshift, (B) stellar mass and (C) stellar-to-gas mass ratio for individual model galaxies. We plot three galaxies with final stellar mass log$(M_\ast/M_\odot) \sim 9.5, 10, 10.5$ at each of the four epochs where we observe the MZ relation.}
\label{fig:model}
\end{figure*}

The metallicity relation plotted in Figure \ref{fig:zfg} is a significant observational result. {In Section 5 we interpret the relation plotted in Figure \ref{fig:zfg} as a redshift independent relation between metallicity and the stellar-to-gas mass ratio, $M_\ast/M_g$. The redshift independence of the relation suggests that it is a universal relation.} The MZ relation is a snapshot of the chemical properties of a population of galaxies at an instant in time. If indeed the relation between metallicity and stellar-to-gas mass ratio is universal, it should apply to individual galaxies as they evolve. Our numerical model allows us to examine the chemical evolution of one galaxy over time. In Figure \ref{fig:model}A, \ref{fig:model}B and \ref{fig:model}C we plot the chemical evolution of individual model galaxies as a function of redshift, stellar mass and stellar-to-gas mass ratio, respectively. We plot three galaxies with log$(M_\ast/M_\odot) \sim 9.5, 10, 10.5$ as they evolve to each of the four epochs where we observe the MZ relation. Each galaxy in this plot finishes on the model MZ relation plotted in Figure \ref{fig:model_fit}, i.e. the end points of the evolution in Figure \ref{fig:model}B are the MZ relations at the four epochs. Figure \ref{fig:model}C shows galaxies evolve along a universal relation between metallicity and stellar-to-gas mass ratio. The results of our numerical modeling support our interpretation based on analytical models.

\section{The Observed Relation Between Metallicity and the Stellar-to-Gas Mass Ratio}

We can observationally test our interpretation of the universal metallicity relation presented in Figure \ref{fig:zfg} with the ALFALFA sample cross-matched with SDSS (see Section 2.5). 

We determine metallicities from strong nebular emission lines. The metallicity derived from an integrated spectrum is a global nebular luminosity-weighted measurement. Galaxies have gas disks that extend well beyond their stellar and star-forming disks \citep{vanderKruit2011}. The 3.5 arc-minute beam of the ALFALFA covers the outer disk of galaxies in the cross-matched sample. To make a proper apples-to-apples comparison of the gas mass and metallicity, we need to derive a gas mass that is weighted in a manner comparable to the nebular luminosity-weighting of the metallicity. The nebular luminosity is dominated by young, massive stars. The luminosity of young massive stars is proportional to the SFR. {In order to properly compare gas mass measurements with the luminosity-weighted metallicity we measure, we derive the SFR-weighted hydrogen gas mass by applying an average weighting to individual measurements of gas mass.}

\citet{Bigiel2012} measure the hydrogen gas (HI + H$_2$) profile using resolved measurements from the HERACLES \citep{Leroy2009} and THINGS \citep{Walter2008} surveys. They observe a universal gas profile when the scale length is normalized to $R25$ and the surface density is normalized to the surface density at the transition radius where the HI to H$_2$ gas masses are equal. $R25$ is defined as the 25 mag arcsec$^{-2}$ B-band isophote. Within $R25$ the measured gas profiles exhibits $\sim25 - 40\%$ scatter. The universal gas profile \citeauthor{Bigiel2012} derive is
\begin{equation}
\Sigma_{\mathrm{gas}} \propto  e^{-1.65r/r_{25}}.
\end{equation}
Here, $r/r_{25}$ is the radius in units of $R25$. \citet{Leroy2008} combine FUV \emph{GALEX} and 24 $\mu$m \emph{Spitzer} imaging to measure the SFR profile. We calculate the average SFR profile in units of $R25$ from their measurements of 18 galaxies with $M_\ast > 10^{9} M_\odot$. The average SFR profile is 
\begin{equation}
\Sigma_{\mathrm{SFR}} \propto  e^{-4.34r/r_{25}} 
\end{equation}
There is $\sim30\%$ scatter in the average profile that we calculate.

We determine the SFR-weighted hydrogen gas mass from the gas and star formation rate profiles. The weighting factor is
\begin{equation}
W_{\mathrm{SFR}} = \frac{\int_{0}^\infty r \, \Sigma_{\mathrm{gas}} \, \Sigma_{\mathrm{SFR}} \, \mathrm{d}r }{\int_{0}^\infty r \, \Sigma_{\mathrm{gas}} \, \mathrm{d}r \int_{0}^\infty  \Sigma_{\mathrm{SFR}} \, \mathrm{d}r } = 0.33.
\end{equation}
On average, 1/3 of the gas in galaxies is located within their star-forming disks. The SFR-weighted hydrogen gas mass is 
\begin{equation}
M_{\mathrm{H}} = W_{\mathrm{SFR}} \, M_{\mathrm{H, measured}}.
\end{equation}
Here $M_{\mathrm{H, measured}}$ is the measurement from ALFALFA with the H$_2$ contribution estimated using scaling relations presented in \citet[see Section 2.5]{Saintonge2011}. The combined scatter from the gas and SFR profiles is $\sim0.3$ dex. This lower limit for the uncertainty for the SFR-weighted hydrogen mass measurement does not include the observational uncertainty associated with the measurement.


\begin{figure}
\begin{center}
\includegraphics[width=\columnwidth]{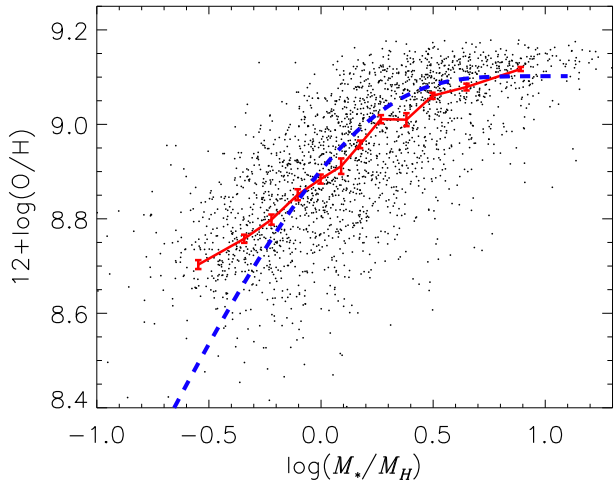}
\end{center}
\caption{The metallicity as a function of the stellar-to-gas mass ratio. The black points are 2633 individual galaxies where we are able to estimate the metallicities, stellar and gas masses from observations. The red curve is the median metallicity in 12 equally populated bins of stellar-to-gas mass ratio. The blue curve is derived by combining Equation \ref{eq:mzfit} and \ref{eq:momg} and adopting the locally measured value of $Z_o = 9.102$.}
\label{fig:mgz}
\end{figure}

In Figure \ref{fig:mgz} we observationally test our interpretation of the universal metallicity relation by plotting the metallicity as a function of the stellar-to-gas mass ratio observed in local galaxies. The black points are the 2633 galaxies in our cross-matched sample. The stellar masses, metallicities and hydrogen gas masses are directly measured. The blue dashed curve is the universal relation between metallicity and stellar-to-gas mass ratio plotted in Figure \ref{fig:zfg} using the relation given by Equation \ref{eq:momg}. The median metallicity in 12 equally populated bins of stellar-to-gas mass ratio is plotted by the red curve. The metallicity errors are bootstrapped. The stellar masses have $\sim 0.1$ dex observational uncertainty. This combined with the $>0.3$ dex uncertainty in the hydrogen mass estimates means that the error on an individual measurement of the stellar-to-gas mass ratio is at least $0.4$ dex. The blue and red curves differ by $\lesssim0.1$ dex. This is good agreement in light of the large uncertainties involved in this measurement. We conclude that the observational data are consistent with the universal relation of chemical evolution.

\section{Discussion}

We analyze observations of the MZ relation for $z\lesssim1.6$. The MZ relation exhibits very simple redshift evolution which we parameterize by redshift evolution in the characteristic turnover mass where the metallicity begins to saturate. We physically interpret our fit parameters by comparing them to solutions of analytical models of chemical evolution. We test and strengthen our interpretation by numerically modeling the chemical evolution of individual galaxies. We observationally test our interpretation and show that it is consistent with the best data currently available. Our analysis provides a simple and intuitive perspective of galactic chemical evolution. \emph{Galaxies follow a universal relation between metallicity and stellar-to-gas mass ratio as they evolve}. Metallicity is defined as the ratio of oxygen to hydrogen. The net mass of oxygen produced is directly proportional to galaxy stellar mass. Given the definition of metallicity, a universal relation between metallicity and stellar-to-gas mass ratio should be expected.

\begin{figure}
\begin{center}
\includegraphics[width=\columnwidth]{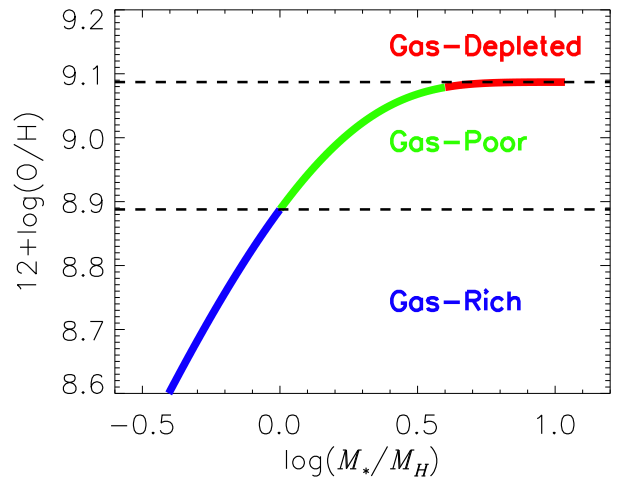}
\end{center}
\caption{A schematic illustrating the three regimes of galactic chemical evolution.}
\label{fig:cartoon}
\end{figure}

The chemical evolution of galaxies can be characterized as having three distinct regimes: gas-rich, gas-poor and gas-depleted. Figure \ref{fig:cartoon} is a schematic which illustrates the three regimes. The gas-rich regime is plotted in blue. Galaxies are in the gas-rich regime when $M_g>M_\ast$. In the gas-rich regime, the metallicity is proportional to stellar-to-gas mass ratio. This can be seen by Taylor expanding Equation \ref{eq:mzr}. The gas-poor regime is plotted in green. Galaxies cross over to the gas-poor regime when $M_g<M_\ast$. In the gas-poor regime the metallicity is high enough that the mass of oxygen that is being locked up in stars becomes an appreciable fraction of the mass of oxygen produced. Galaxies exponentially approach the metallicity saturation limit. The gas-depleted regime is plotted in red. Galaxies in the gas-depleted regime have $M_g/M_\ast \ll 1$. In this regime, the metallicity is so high that the mass of oxygen taken out of the ISM and locked up in low mass stars equals the mass of oxygen produced by massive stars. The metallicity can not increase beyond this point.

The MZ relation originates from the more fundamental universal relation between metallicity and stellar-to-gas mass ratio. At a fixed stellar mass, the metallicities of galaxies increase as the Universe evolves because of a commensurate decline in their gas content. The slope of the MZ relation at $M_\ast < M_o$ is set by the slope of the relation between gas mass and stellar mass. This is because the slope of the more fundamental relation between metallicity and gas-to-stellar mass ratio is unity for galaxies with $M_\ast < M_o$ (see Equation \ref{eq:mzr}). The scatter observed in the MZ relation largely reflects the scatter in gas mass at a fixed stellar mass. The relation between stellar mass, metallicity and SFR \citep[e.g.][]{Mannucci2010} is a natural consequence of the universal metallicity relation. At a fixed stellar mass, galaxies with larger gas reservoirs will typically exhibit elevated SFRs and diluted metallicities and vice versa. Several authors have recognized variations in the gas content of galaxies as the basis of the observed relation between stellar mass, metallicity and SFR \citep[e.g.,][]{Dave2012, Dayal2013, Bothwell2013, Lilly2013, Dave2013}. Our results are consistent with this interpretation.

The universal relation of chemical evolution implies that the metallicity of a galaxy is instantaneously set by its stellar-to-gas mass ratio. Gas flows and star formation move galaxies along$- \emph{not off} -$the universal metallicity relation. Unlike the MZ relation in which the stellar mass is monotonically increasing, the stellar-to-gas mass ratio can increase or decrease. The response of galaxies to large gas accretion events will be to instantaneously move down the universal metallicity relation. Conversely, galaxies will move up the universal metallicity relation as they deplete their gas reservoirs by star formation and outflows. 

The intrinsic scatter in the universal metallicity relation is likely to be small. At a fixed stellar mass we expect some scatter in the net yield, $Y_N$ due to scatter in total mass of oxygen expelled from galaxies. Any scatter in the net yield would directly translate into scatter in the universal metallicity relation. Mergers and starbursts will likely also contribute to the scatter in the universal metallicity relation since these events significantly disrupt the ISM of galaxies. The observed scatter in the FMR of \citet{Mannucci2010} and HI FMR of \citet{Bothwell2013} is $\sim0.07$ dex. Since the FMR is directly a result of the universal metallicity relation, the measured scatter in the FMR provides an upper limit for the scatter in the universal metallicity relation.

If our interpretation is correct, the metallicities of galaxies provide a precise probe of their ISM gas content. We derive the relation between average gas mass and stellar mass by combining Equations \ref{eq:mo} and \ref{eq:momg}. The average mass of hydrogen gas in the star-forming disk of galaxies as a function of redshift and stellar mass is
\begin{equation}
M_g(M_\ast, z) = 3.87\times10^9 \, (1+z)^{1.35} \left(\frac{M_\ast}{10^{10}M_\odot}\right)^{0.49} \, \, [M_\odot].
\label{eq:gas_redshift}
\end{equation}
In the absence of direct measurements, the gas mass of galaxies can be estimated from measurements of stellar mass and metallicity by combining Equation \ref{eq:mzfit} and \ref{eq:momg} and solving for $M_g$. The hydrogen gas mass as a function of stellar mass and metallicity is
\begin{equation}
M_g(M_\ast, Z) = \frac{-M_\ast}{\mathrm{ln}\left(1 - 10^{\left[Z - Z_o\right]} \right)} \, \, [M_\odot].
\label{eq:gas_metallicity}
\end{equation}
The stellar mass is measured in units of $M_\odot$ and $Z_o = 9.10$. We emphasize that metallicities must be converted into the KK04 diagnostic when using Equation \ref{eq:gas_metallicity}. Gas masses estimated using Equation \ref{eq:gas_metallicity} carry the uncertainties associated with stellar masses, metallicities and the intrinsic scatter in the universal metallicity relation. {The KK04 strong-line metallicity method is estimated to have an accuracy of $\sim0.15$ dex \citep{Kewley2008} and we estimate that stellar masses are accurate to $\sim0.1$ dex once systematic differences between methods are removed (see Section 3.1). Combining these uncertainties with the $\lesssim0.07$ dex estimated scatter in the universal metallicity relation, we conclude that the gas masses estimated using Equation \ref{eq:gas_metallicity} are accurate to within $\sim0.3$ dex.} The relations presented by Equation \ref{eq:gas_redshift} and \ref{eq:gas_metallicity} are valid for galaxies at $z\lesssim1.6$. These relations are testable predictions of the gas content of galaxies based on the observed evolution of the MZ relation. However, we note that in the saturation regime, the metallicity is not sensitive to the stellar-to-gas mass ratio.

A tight relation between stellar mass and SFR (MS relation) is observed to extend out to at least $z\sim2$ \citep{Noeske2007a, Salim2007, Elbaz2007, Pannella2009, Whitaker2012, Zahid2012b, Kashino2013}. The slope of the MS relation does not evolve significantly with redshift and the zero point declines by a factor of $\sim20$ since $z\sim2$. The slope of the MZ relation is set by the slope of the relation between gas mass and stellar mass. From our observations we conclude that the slope does not evolve significantly for $z\lesssim1.6$. The nearly constant slope of the MS relation is consistent with a constant slope that we infer for the relation between gas mass and stellar mass based on our measured MZ relations. From Equation \ref{eq:gas_redshift}, we estimate that the zero point of the relation between gas mass and stellar mass evolves by a factor of $\sim3$ since $z\sim1.6$. This evolution is significantly smaller than the zero point evolution of the MS relation. This implies that the star formation efficiency, the SFR relative to gas mass, increases with redshift. This increase may be the result of higher molecular-to-atomic hydrogen mass ratios in high redshift galaxies \citep[see][]{Dutton2010}.

We have provided evidence for a universal relation between metallicity and stellar-to-gas mass ratio for galaxies at $z\lesssim1.6$. Our observations only extend down to stellar masses of $\sim10^{9} M_\odot$ and gas-to-stellar mass ratios of $\sim0.5$. By summing the spectra of $\sim200,000$ galaxies in the SDSS, \citet{Andrews2013} measure the MZ relation down to stellar masses of $\sim10^{7.5} M_\odot$. They measure a continuous MZ relation which flattens at high stellar masses and scales as $O/H \propto M_\ast^{1/2}$ at stellar masses $<10^{9} M_\odot$. Our measurements are consistent with the scaling of the MZ relation measured by \citet{Andrews2013} suggesting that the universal metallicity relation may extend well into the dwarf regime.

Additional observational tests are needed to establish the universal metallicity relation. Measurements of stellar masses, metallicities and gas masses of dwarf galaxies will allow us to directly measure how low in stellar mass and stellar-to-gas mass ratio the universal metallicity relation extends . Metal-rich dwarf galaxies have been observed \citep{Zahid2012a}. These galaxies will provide an interesting test of the university metallicity relation. Measurements of gas masses for galaxies outside the local Universe will provide the most important test of the universal metallicity relation. The combination of \emph{ALMA} and \emph{SKA} will soon make this possible. Finally, we have examined the global properties of galaxies. The universal relation should be tested in galaxies where spatially resolved measurements of the metallicity and gas content are available. {Indeed, \citet{Ascasibar2014} report a fundamental relation between stellar-to-gas mass ratio and metallicity that holds on $\sim1$ kpc scales.} Moreover, several integral field surveys of nearby galaxies currently underway (e.g., SAMI, WALLABY, MANGA) should prove useful for further investigating the spatially resolved nature of the universal metallicity relation.

The physical basis of the universal metallicity relation needs to be explored in much greater detail. Several aspects of this work imply a cosmological origin for the relation. The evolution of the characteristic turnover mass, $M_o$, is proportional to $(1+z)^{2.6}$. This scaling with redshift is similar to the $(1+z)^{2.5}$ scaling of the growth rate of dark matter halos in simulations \citep{Fakhouri2010}. Additionally, our observations suggest that $M_g \appropto \sqrt{M_\ast}$ for $z\lesssim1.6$. This scaling of gas and stellar mass is similar to the scaling between stellar and halo mass, $M_h$, determined from abundance matching \citep{Behroozi2013} and from galaxy-galaxy weak lensing, galaxy clustering and galaxy distribution \citep{Leauthaud2012}. This implies that $M_g/M_h$ is nearly constant with respect to stellar mass. The slope of the relation between stellar mass and halo mass does not evolve significantly out to $z\sim1.6$ \citep{Leauthaud2012, Behroozi2013}. An $M_g/M_h$ ratio that is constant with stellar mass may be the physical origin for the constant slope of the relation between gas mass and stellar mass. 

A detailed analysis of systematic issues related to measurements of the MZ relation should soon be possible as larger data sets become available. In particular, metallicities are derived using techniques that are calibrated using observations of local galaxies. If ISM conditions evolve with redshift, these calibrations may need to be revised. Our analysis is primarily based on the R23 diagnostic calibrated by \citet{Kobulnicky2004}. However, long-standing discrepancies between various metallicity diagnostics remain unresolved \citep[e.g.,][]{Kudritzki2008, Bresolin2009b, Kudritzki2012, Nicholls2012, Dopita2013}. Discrepancies between various metallicity diagnostics do not change the major conclusions of this work but the quantitative analysis presented is subject to these uncertainties. Additionally, larger and more complete spectroscopic samples combined with cosmological simulations should allow us to assess selection biases in our measurement of the MZ relation. The MZ relation appears to be insensitive to selection biases and we anticipate that correction for these biases will not change the major conclusions presented in this work.


\section{Summary and Conclusions}

We measure the MZ relation for $z\lesssim1.6$ using a consistent methodology. We interpret our observations with the aid of analytical and numerical models. We propose a new paradigm for understanding the chemical evolution of galaxies. The main results of our analysis are that:
\begin{itemize}

\item  The evolution of the MZ relation for $z\lesssim1.6$ is very simple. The evolution can be parameterized solely by the redshift dependency of the  characteristic turnover mass, $M_o$. $M_o$ is the stellar mass where the MZ relation begins to flatten.The relation between metallicity and stellar mass normalized to $M_o$ is independent of redshift for $z\lesssim1.6$. 

\item We physically interpret the parameters we fit to the MZ relation using analytical and numerical models. The MZ relation saturates at a maximum metallicity, $Z_o$. The saturation occurs when the gas-phase abundance is high enough that the mass of oxygen locked up in low mass stars equals the mass of oxygen produced by massive stars. Galaxies are not able to enrich beyond this metallicity. The slope of the MZ relation, $\gamma$, is set by the slope of the relation between gas mass and stellar mass. The characteristic stellar mass, $M_o$, is the stellar mass where the stellar-to-gas mass ratio is unity. 

\item  We show that the redshift independent metallicity relation is a relation between metallicity and stellar-to-gas mass ratio. Numerical modeling suggests that all galaxies follow this metallicity relation as they evolve. We refer to the relation between metallicity and stellar-to-gas mass ratio as the universal metallicity relation. 

\item The MZ relation originates from the universal metallicity relation. The evolution of the MZ relation is due to the evolving gas content of galaxies.

\item  We directly measure the stellar masses, metallicities and gas masses for a sample of local galaxies. These data are consistent with the universal metallicity relation.

\item  We show that the chemical evolution of galaxies can be characterized by three distinct regimes of evolution. In the gas-rich regime when $M_g > M_\ast$ the metallicity is proportional to the stellar-to-gas mass ratio. In the gas-poor regime when $M_g < M_\ast$ metallicity exponentially approaches the saturation limit. In the gas-depleted regime when $M_g \ll M_\ast$, the metallicity is saturated and does not increase beyond this limit.

\item  The observed evolution of the MZ relation is due to the evolving of gas content of galaxies. We derive the average gas mass as function of stellar mass and redshift. The universal metallicity relation is a relation between stellar mass, metallicity and gas mass. We provide an equation to estimate gas mass from measurements of stellar mass and metallicity.

\end{itemize}
We outline some observational tests necessary to further validate the universal metallicity relation. \emph{ALMA} combined with \emph{SKA} will be transformative for our understanding of the gas content of galaxies. These facilities should allow us to directly test the universal metallicity relation out to high redshifts.

\acknowledgements
We thank Molly Peeples, Paul Torrey, Manolis Papastergis and Charlie Conroy for useful discussions contributing to this work. This work was supported by the National Science Foundation under grant AST-1008798 to R.P.K. The authors wish to recognize and acknowledge the very significant cultural role and reverence that the summit of Mauna Kea has always had within the indigenous Hawaiian community.  We are most fortunate to have the opportunity to conduct observations from this mountain. Mahalo.

Some of the data reported here were taken with the Subaru telescope as part of the FMOS-COSMOS survey. We thank the COSMOS collaboration.
 
Some of the data reported here were obtained at the MMT Observatory, a joint facility of the University of Arizona and the Smithsonian Institution.

Funding for the SDSS and SDSS-II has been provided by the Alfred P. Sloan Foundation, the Participating Institutions, the National Science Foundation, the U.S. Department of Energy, the National Aeronautics and Space Administration, the Japanese Monbukagakusho, the Max Planck Society, and the Higher Education Funding Council for England. The SDSS Web Site is http://www.sdss.org/.

The SDSS is managed by the Astrophysical Research Consortium for the Participating Institutions. The Participating Institutions are the American Museum of Natural History, Astrophysical Institute Potsdam, University of Basel, University of Cambridge, Case Western Reserve University, University of Chicago, Drexel University, Fermilab, the Institute for Advanced Study, the Japan Participation Group, Johns Hopkins University, the Joint Institute for Nuclear Astrophysics, the Kavli Institute for Particle Astrophysics and Cosmology, the Korean Scientist Group, the Chinese Academy of Sciences (LAMOST), Los Alamos National Laboratory, the Max-Planck-Institute for Astronomy (MPIA), the Max-Planck-Institute for Astrophysics (MPA), New Mexico State University, Ohio State University, University of Pittsburgh, University of Portsmouth, Princeton University, the United States Naval Observatory, and the University of Washington."

Some of the data presented herein were obtained at the W.M. Keck Observatory, which is operated as a scientific partnership among the California Institute of Technology, the University of California and the National Aeronautics and Space Administration. The Observatory was made possible by the generous financial support of the W.M. Keck Foundation.

Funding for the DEEP2 Galaxy Redshift Survey has been provided by NSF grants AST-95-09298, AST-0071048, AST-0507428, and AST-0507483 as well as NASA LTSA grant NNG04GC89G.

\bibliographystyle{apj}
\bibliography{/Users/jabran/Documents/latex/metallicity}

 \end{document}